\documentclass[fleqn,usenatbib]{mnras}
\usepackage{newtxtext,newtxmath}

\usepackage{graphicx}	
\usepackage{amsmath}	
\usepackage{amssymb}	
\usepackage{tabularx}
\usepackage{subfigure}
\usepackage[flushleft]{threeparttable}
\usepackage{hyperref}
\usepackage{gensymb}

\newcommand{\msun}{M_\odot}

\newcommand{\gcm}{\rm ~g~cm^{-3}}
\newcommand{\kevcm}{\rm ~keV~cm^{-3}}
\newcommand{\cmc}{\rm ~cm^{-3}}
\newcommand{\ergs}{\rm ~erg~s^{-1}}
\newcommand{\Myr}{\rm ~Myr}
\newcommand{\Rch}{R_{\rm ch}}
\newcommand{\thetah}{\theta_{\rm h}}
\newcommand{\tb}{t_{\rm b}}
\newcommand{\tage}{t_{\rm age}}
\newcommand{\tchoke}{t_{\rm ch}}
\newcommand{\tsp}{t_{\rm sp}}
\newcommand{\tbuoy}{t_{\rm buoy}}
\newcommand{\tbo}{t_{\rm bo}}
\newcommand{\Rs}{R_{\rm s}}
\newcommand{\Rc}{R_{\rm c}}
\newcommand{\rhoa}{\rho_{\rm a}}
\newcommand{\rhoch}{\rho_{\rm ch}}
\newcommand{\Pa}{P_{\rm a}}
\newcommand{\Pc}{P_{\rm c}}
\newcommand{\Lj}{L_{\rm j}}
\newcommand{\Ns}{N_{\rm s}}
\newcommand{\ntot}{n_{\rm t}}
\newcommand{\Rcore}{R_{\rm core}}
\newcommand{\Mcore}{M_{\rm core}}
\newcommand{\vt}{v_{\rm t}}
\newcommand{\vk}{v_{\rm k}}
\newcommand{\betah}{\beta_{\rm h}}
\newcommand{\betac}{\beta_{\rm c}}
\newcommand{\Vfs}{V_{\rm fs,ch}}
\newcommand{\Pch}{P_{\rm c,ch}}
\newcommand{\gammaj}{\gamma_{\rm j}}
\newcommand{\Sigmah}{\Sigma_{\rm h}}
\newcommand{\teqlat}{t_{\rm eq,lat}}
\newcommand{\zh}{z_{\rm h}}

\title[Jet-driven bubbles]{Jet-driven bubbles in Fanaroff--Riley type I sources}

\author[C. M. Irwin et al.]{
Christopher M. Irwin,$^{1}$\thanks{E-mail: christopheri@mail.tau.ac.il} 
Xiaping Tang,$^{2}$ 
Tsvi Piran$^{2}$
and Ehud Nakar$^{1}$
\\
$^{1}$The Raymond and Beverly Sackler School of Physics and Astronomy, Tel Aviv University, Tel Aviv 69978, Israel \\
$^{2}$The Racah Institute of Physics, The Hebrew University of Jerusalem, Jerusalem 91904, Israel
}

\date{Accepted XXX. Received YYY; in original form ZZZ}

\pubyear{2019}

\begin{document}
\label{firstpage}
\pagerange{\pageref{firstpage}--\pageref{lastpage}}
\maketitle

\begin{abstract}
Observations of several Fanaroff--Riley (FR) type I sources reveal outflowing bipolar bubbles of hot gas surrounded by a weak forward shock. We consider the possibility that these bubbles were driven by choked relativistic jets which failed to penetrate the ambient intracluster medium (ICM).  Using new results on choked jets linking the geometry of the forward shock to the jet properties, we infer robust limits on the radius $\Rch$ at which the jet was quenched in 5 well-studied FRI sources, finding typically $\Rch\sim 10$\,kpc.   We further show that, in order to reach this radius in less than the current age of the system, the jet must have been tightly collimated, with the jet head subtending an angle of $\thetah \la 2\degree$.  The ambient pressure is not high enough to explain this collimation, suggesting that the jet was collimated by interaction with its own cocoon.  Although the choking radius is well-constrained, we find a degeneracy between the initial jet opening angle before collimation, $\theta_0$, and the duration of jet activity, $\tb$, with $(\tb/1\Myr)(\theta_0/5\degree)^{-2}\sim0.1$.  We speculate that the working time and/or opening angle of the jet may be important factors contributing to the FR type I/type II morphology in galaxy clusters, with short-lived or wide jets being choked to form bipolar bubbles filled with diffuse radio emission, and longer-lived or narrow jets successfully escaping the cluster core to produce cocoons with radio hotspots.
\end{abstract}


\section{Introduction} \label{sec:introduction}

With high-resolution X-ray imaging, the Chandra and XMM-Newton space telescopes have identified many cavities with a depression of surface brightness in galaxies, galaxy groups, and galaxy clusters \citep[e.g.,][]{Birzan04,Diehl08,DF06,Dong10,Shin16}. The X-ray cavities are often filled with radio lobes and are thought to be inflated by active galactic nuclei (AGN) jets \citep[e.g.,][]{Churazov00,McNamara00,Fabian02,McNamara05}. Weak shocks surrounding the cavity are also observed in several well-studied objects like the Perseus cluster and M87 \citep[e.g.,][]{Graham08,Forman17}, which further supports the connection with jet activity. The jet energy required to produce the X-ray cavities can balance the cooling of the hot gas \citep[e.g.,][]{Birzan04,Rafferty06}, which becomes an important sign of radio mode AGN feedback; see, e.g., \cite{Fabian12} for a recent review.

Some Fanaroff--Riley (FR) type I sources \citep{FR74} which contain X-ray cavities also show clear evidence for bipolar relativistic jets.  Examples include M87 \citep[e.g.,][]{Hines89,Owen00,Marshall02,Forman07,Forman17} and Perseus \citep[e.g.,][]{Boehringer93,Fabian00,Churazov00}.  It is therefore natural to consider whether the bipolar bubbles we observe could be relics of past jet activity.  If this is the case, what can we learn about the jet by observing the relic bubbles it leaves behind?  Thanks to the wealth of observational data available, it is often possible to obtain robust estimates for the age ($\tage$) of the bubbles,\footnote{We define $\tage$ as the time since the jet was launched.} as well as the total energy ($E$) injected into the system \citep[][see also Section~\ref{sec:model}]{TC17}.  However, the properties of the engine responsible for driving the outflow are more uncertain.  For example, we do not know whether energy injection is still ongoing, or ceased long ago.  Likewise, it is not clear whether the energy was injected quasi-spherically, or concentrated into a narrow angle.

Motivated by these questions, we investigate a choked relativistic jet model for the bubbles in FR type I objects.  In the choked jet scenario, we assume that the bubbles were driven by a relativistic jet that was quenched some time in the past, so that we no longer observe it (see Sections~\ref{sec:jet} and~\ref{sec:choking}).  We demonstrate how four observed bubble properties (the radius of the forward shock, $\Rs$; the location of the contact discontinuity, $\Rc$; the shock Mach number, $M$; and the bubble's apparent aspect ratio) can be used to constrain the four parameters of our model (the total injected energy, $E$; the age of the system, $\tage$; the duration of the jet outburst, $\tb$; and the initial jet opening angle, $\theta_0$).\footnote{To help keep track of the meanings of symbols introduced throughout the paper, we provide a glossary in Table~\ref{table:symbols}.}

The possibility that the bubbles in FR I sources were driven by quenched outbursts has been studied analytically before by, e.g., \cite{TC17} and references therein.   Our model differs from theirs in two key ways.  First, whereas they assume quasi-spherical energy injection, we assume that the energy was injected by a jet into a nozzle of opening angle $\theta_0$.  Second, we make use of an additional observable, the apparent aspect ratio of the forward shock.  In other words, our model has one additional free parameter, and one additional observational constraint.  

In applying our model, we focus on 5 FR type I sources--the Perseus cluster \citep{Zhuravleva16}, M87 \citep{Forman17}, MS 0735.6+7421 \citep{Van14}, NGC 4552 \citep{Mach06}, and NGC 5813 \citep{Randall15}--which have a pair of bubbles close to the central AGN surrounded by a clearly visible forward shock.  (See, e.g., Figure 6 of \citet{Zhuravleva16} for an X-ray image of the relevant features.)  In these cases, $\Rs$, $\Rc$, $M$, and the shock aspect ratio are all reliably measured.  Each object has a low Mach number ($M<2$) and $\Rs \sim \Rc$, although the size of the systems varies across two orders of magnitude, from $\Rs \sim 3$\,kpc in NGC 4552, to $\Rs \sim 300$\,kpc in MS 0735.6+7421.  In all but one of the systems, the apparent aspect ratio of the forward shock is $\sim 1.5$, and the projected distance between the black hole and the centre of each bubble is comparable to the bubble's size.  (M87 is the exception; see Section~\ref{sec:discussion} for further discussion.)  The ambient gas density ($\rhoa$) and pressure ($\Pa$) in these environments are also well-constrained, with typical values of $\rhoa \sim 10^{-25} \gcm$ and $\Pa \sim 0.1 \kevcm$.  In addition, we make use of the density profile power-law index ($\alpha$) reported in the ACCEPT database \citep{Cava09}.  The values of $\alpha$ range from 0 to 1.5.   All the observational data are summarised in Table~\ref{table:data}. 

We begin our discussion with an overview of the evolution of the system, first considering the behaviour while the central engine is active in Section~\ref{sec:jet}, and then addressing what happens once the jet switches off in Section~\ref{sec:bubbles}.  We review several important time-scales for bubble evolution in Section~\ref{sec:timescales}, then show how available observations of relic bubbles constrain the jet geometry (Section~\ref{sec:choking}) and the duration of engine activity (Section~\ref{sec:model}).  As the specific objects we consider are all located in clusters, we focus here on the propagation in cluster environments.  However, much of the discussion is generic and applies also to isolated radio galaxies.  In Section~\ref{sec:discussion}, we present our results for each individual object, and consider possible implications for the FRI/FRII dichotomy, before concluding in Section~\ref{sec:conclusion}.

\section{Jet propagation and collimation}\label{sec:jet}

A relativistic AGN jet drives a strong forward shock into the ambient medium, while at the same time a reverse shock is driven back into the jet ejecta.  The resulting double shock structure is known as the ``jet head.''  As the jet propagates,  jet material entering the head through the reverse shock is pushed out to the sides, forming a hot ``cocoon'' that surrounds the jet.  The subsequent evolution of the jet-and-cocoon system depends on the relationship between the thermal pressure in the cocoon, $\Pc$, and the sum of all pressures in the ambient medium at the jet head's location.  We suppose that the ambient density and pressure are power-laws in radius, with $\rho \propto R^{-\alpha}$ and $P \propto R^{-\lambda}$, and denote the ambient density and pressure at the location of the forward shock along the axis as $\rhoa$ and $\Pa$, respectively.  At any given time, the relevant behavior is then determined by comparing the values of $\Pc$ and $\Pa$.

If $\Pc \gg \Pa$, the cocoon expands supersonically, driving a strong shock into the external medium; we call this shock the ``cocoon shock.''  At the same time, the cocoon exerts pressure on the conical jet outflow, reducing its opening angle.  A self-consistent analytical solution for the dynamics of the jet and cocoon in this regime is given by \citet[hereafter B11]{BNPS11}. The resulting evolution depends on the ambient density $\rhoa$, as well as the jet's luminosity $\Lj$ and injection angle $\theta_0$, but it is independent of the ambient pressure.  

\begin{figure}
\begin{center}
\includegraphics[width=\columnwidth]{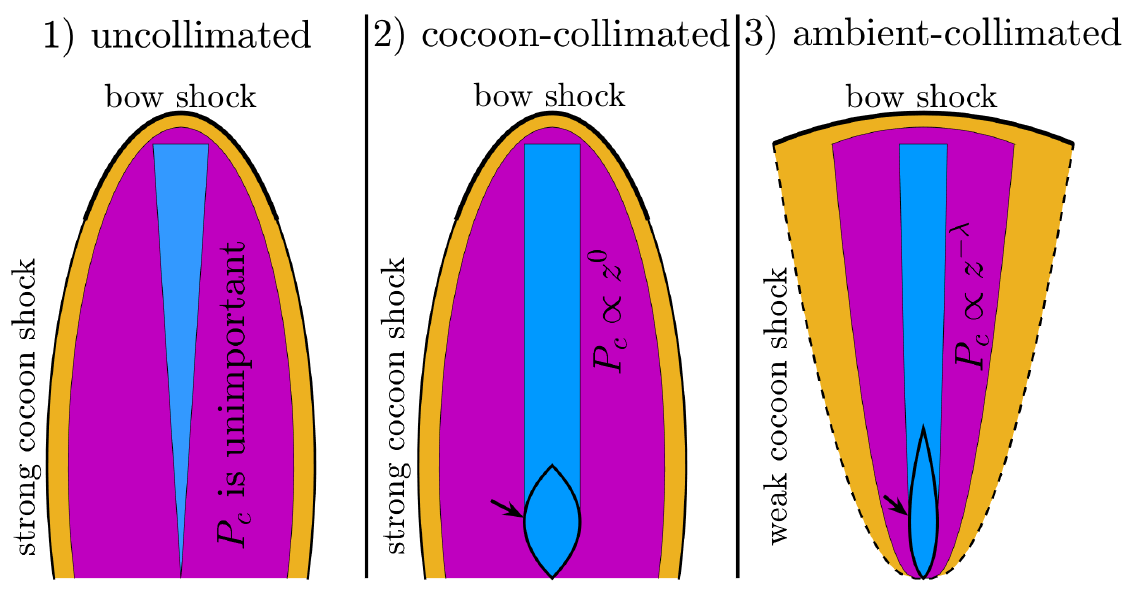} 
\caption{Possible regimes of jet collimation.  The jet, cocoon, and shocked ambient medium are coloured blue, purple, and yellow respectively.  The collimation shock is indicated with an arrow.  In the cocoon-collimated case the cocoon pressure $P_c$ is spatially uniform, but in the ambient-collimated case it is a power-law in the height $z$.}
\label{fig:collimation}
\end{center}
\end{figure}

The $\Pc \gg \Pa$ regime can be further subdivided based on the ratio of the jet energy density at the head, $\Lj/\Sigmah c$ (where $\Sigmah$ is the cross section of the jet head and $c$ is the speed of light), to the ambient rest-mass energy density, $\rhoa c^2$.   Following B11, we call this dimensionless ratio $\tilde{L} \equiv \Lj/(\Sigmah \rhoa c^3)$.  As $\tilde{L}$ increases, the pressure in the cocoon grows, but the pressure required to collimate the jet grows even faster.  Consequently, there is a critical value of $\tilde{L}$ (or, equivalently, a critical value of $\Pc$) above which the cocoon pressure is no longer sufficient to collimate the jet.  
If $\tilde{L} \gg \theta_0^{-4/3}$ (or $\Pc \gg \rhoa c^2 \theta_0^{2/3}$), the cocoon pressure is not high enough to significantly alter the jet opening angle.  (Note that, for a typical jet opening angle of $\sim$ several degrees and an ICM temperature of $\sim$ a few keV, $\rhoa c^2 \theta_0^{2/3} \gg \Pa$.)  In this case, the jet outflow remains conical with an opening angle $\approx \theta_0$, as in the leftmost panel of Fig.~\ref{fig:collimation}.  We refer to this as the ``uncollimated'' regime.   On the other hand, if $\tilde{L} \ll \theta_0^{-4/3}$ (or $\Pc \ll \rhoa c^2 \theta_0^{2/3}$), the cocoon pressure is sufficient  to collimate the jet, transforming it from a conical to a cylindrical flow, as in the middle panel of Fig.~\ref{fig:collimation}.  Deflection of the flow is achieved via an oblique ``collimation shock'' which forms near the base of the jet, as indicated by the heavy black line in the figure.  The cylindrical shape of the jet in this ``cocoon-collimated'' regime is a consequence of the near-uniform pressure within the cocoon (B11).

The situation changes if the pressure in the cocoon becomes similar to the ambient pressure.  In this case, the cocoon is confined by the pressure of the ambient medium, and the sideways expansion stops once pressure balance is achieved.  At a given height, the timescale to reach pressure equilibrium can be estimated by noting that, as long as the jet head is non-relativistic (see Section~\ref{sec:choking}) and external pressure is not yet important, the cocoon pressure scales with time as $\Pc \propto t^{-(4+\alpha)/(5-\alpha)}$ (B11).  Now, suppose that at a height $z$, pressure equilibrium in the lateral direction is achieved at a time $\teqlat(z)$.  The cocoon pressure at the time $\teqlat$ is then related to the current cocoon pressure $\Pc \equiv \Pc(\tage)$ via $\Pc(\teqlat)/\Pc(\tage) =  (\teqlat/\tage)^{-(4+\alpha)/(5-\alpha)}$.  In a similar fashion, since $\Pa\equiv \Pa(\zh)$ is the ambient pressure at the location of the jet head, the ambient pressure as a function of $z$ can be written as $\Pa(z)/\Pa(\zh) = (z/\zh)^{-\lambda}$, where $\zh$ is the current position of the jet head along the axis.  Setting $\Pc(\teqlat(z)) = \Pa(z)$, we obtain the relation
\begin{equation}
\label{eq:teqlat}
\frac{\teqlat(z)}{\tage} = \left[\left(\frac{\Pc}{\Pa}\right) \left(\frac{z}{\zh}\right)^{\lambda}  \right]^{(5-\alpha)/(4+\alpha)}.
\end{equation}
Inspecting eq.~\ref{eq:teqlat}, we see that when $\Pc \gg \Pa$, the time to reach equilibrium is much longer than the age except at $z \ll \zh$, so the ambient pressure does not affect the evolution much.  However, when $\Pc \sim \Pa$, the time to reach equilibrium is comparable to the age of the system at $z \sim \zh$, and is much shorter than the age for $z \ll \zh$.  This implies that when $\Pc$ and $\Pa$ are comparable, the cocoon pressure and ambient pressure are roughly equal along much of the cocoon's length, with the pressure gradient in the cocoon matching that of the ambient medium.  Unlike the cocoon-collimated case where the jet becomes cylindrical, the jet cross-section in this case is a function of height.  The geometry of the jet and the collimation shock when the cocoon pressure decreases as a power law with height is described by, e.g., \citet{MI13} (see their Section 4.2).  This ``ambient-collimated'' regime is illustrated in the right panel of Fig.~\ref{fig:collimation}.

Because $\Pc$, $\Pa$, and $\rhoa$ all change over time, the system may transition between the different collimation regimes as it evolves. Assuming that the jet head advances non-relativistically, the head position evolves as $\zh \propto t^{3/(5-\alpha)}$ (B11).  Accordingly, the ambient density at the head scales as $\rhoa \propto \zh^{-\alpha} \propto t^{-3\alpha/(5-\alpha)}$, while the ambient pressure at the head's location obeys $\Pa \propto \zh^{-\lambda} \propto t^{-3\lambda/(5-\alpha)}$.  Meanwhile, the cocoon pressure falls of as $\Pc \propto t^{-(4+\alpha)/(5-\alpha)}$, as discussed above.  If the density and pressure profiles are relatively flat, as is typical for ICM environments, then $(4+\alpha)$ is larger than both $3\alpha$ and $3\lambda$, and $\Pc$ falls off faster than $\rhoa$ and $\Pa$.  In this case, the jet starts out uncollimated, then goes through a cocoon-collimated phase, and finally ends up in the ambient-collimated regime at late times.

On kiloparsec length scales, the jet is firmly in the collimated regime for typical AGN parameters (B11; see also Section~\ref{sec:choking}), so we do not further consider the case of an uncollimated jet.  However, whether the jet is collimated by the cocoon (as in panel 2 of Fig.~\ref{fig:collimation}) or by the ambient pressure (as in panel 3) is less clear.  AGN bubble observations impose constraints on the jet properties which offer new insight into the collimation process.  We will revisit this discussion in Section~\ref{sec:choking}. 

We stress that the above discussion is only relevant far from the central black hole, where the jet is propagating into the ICM.  The environment close to the central AGN (i.e., within the Bondi radius) is likely considerably different, and this can affect the collimation on scales of $\sim$ tens of parsecs.  Accretion disc winds, in particular, have been shown to impact the jet geometry over these scales \citep[e.g.,][and references therein]{GL16}.  High-resolution observations of M87 \citep{JBL99,AN12} have also revealed changes in the jet opening angle within $\sim 30$\,pc of the central source.  Our model applies to kiloparsec scales where these effects are unimportant.  The injection opening angle $\theta_0$ adopted here refers to the opening angle of the jet when it escapes the central region and starts to interact with the ICM.

\section{Bubble evolution} \label{sec:bubbles}
\begin{figure}
\begin{center}
\includegraphics[width=\columnwidth]{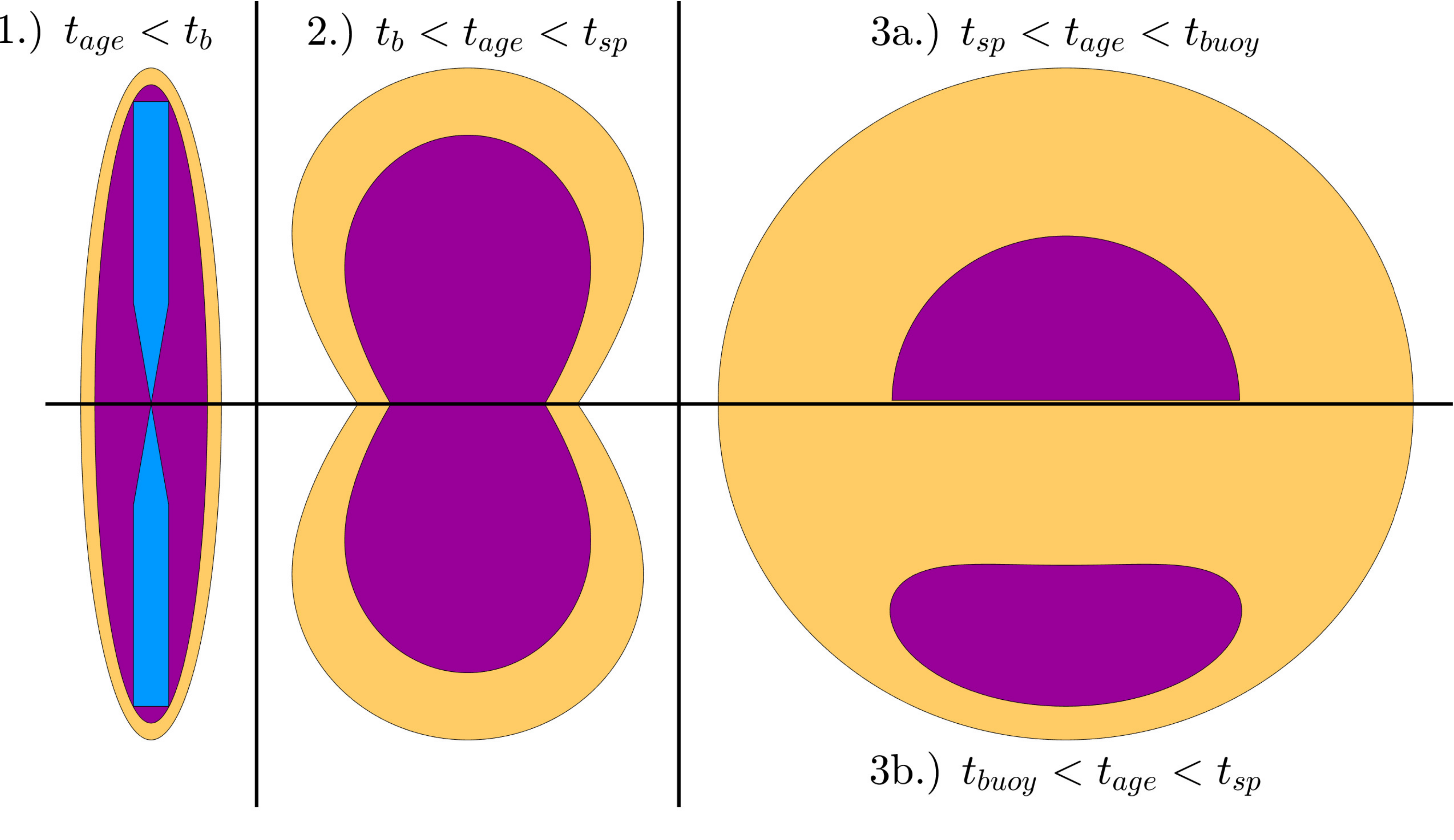} 
\caption{The three phases of evolution for AGN-driven bubbles and shocks. Panel 1) is the jet phase, and panel 2) is the bipolar lobes phase. Panels 3a) and 3b) show the quasi-spherical phase, respectively for the case of no buoyancy (top) and appreciable buoyancy (bottom). Blue is the collimated jet, purple is the cocoon, and yellow is the shocked ambient medium.}
\label{fig:bubbleshape}
\end{center}
\end{figure}

In an outburst, the central AGN launches bipolar jets into the surrounding medium. The subsequent evolution of the system can be divided into three morphological phases, based on the relation between the age of the system ($\tage$), the burst duration ($\tb$), the time it takes for the outflow to become spherical ($\tsp$), and the time-scale for buoyancy to pull the bubbles apart ($\tbuoy$).  These three phases, which we refer to as the jet phase, the bipolar lobes phase and the quasi-spherical phase, are illustrated in Fig. \ref{fig:bubbleshape}.  Two evolutionary paths are possible, depending on whether $\tbuoy < \tsp$ or $\tbuoy > \tsp$.

In the initial jet phase (left panel), the collimated jet (blue) inflates a narrow shocked region along the axis. We assume that the jet has a constant one-sided luminosity $\Lj$, so that the total energy injected by the bipolar jets is $E=2\Lj \tb$.  The purple and the yellow regions represent the shocked jet material and the shocked ambient medium respectively, which are separated by a contact discontinuity.  The X-ray cavities, or ``bubbles,'' seen in observations correspond to the region inside the contact discontinuity, which is filled with light jet exhaust and is ultimately affected by buoyancy.  (For the rest of the paper, we use the terms ``bubble" and ``cocoon" interchangeably, with both referring to the shocked jet ejecta.)

We suppose that the jet which drove the bubbles did not persist for long enough to penetrate the surrounding cluster medium.  Instead, the jet shut off and was choked at a distance $\Rch$ from the central AGN.  (We consider the jet to be ``choked" once all of the jet ejecta have flown through the jet head and entered the cocoon.)  We define the duration $\tb$ as the time when the jet shuts off, and the choking time $\tchoke$ as the time when the last of the jet material catches up to the jet head.  If the jet material below the head is relativistic, $\tb$ and $\tchoke$ are related by 
$\tchoke \approx \tb/(1-\betah)$ \citep{N15},
where $\betah$ is the velocity of the jet head scaled to $c$.  We work in the limit of a Newtonian jet head, with $\betah \ll 1$ (we will verify this assumption in Section~\ref{sec:choking}).  In this case, the non-relativistic jet head does not have time to advance much before the relativistic ejecta catch up, and as a result $\tchoke \approx \tb$.  

In order to estimate the choking radius, we use the results of \citet{HGN18}, who calibrated the analytical expressions of B11 to accurately match numerical simulations.  We assume here that the jet was in the cocoon-collimated regime (this will be verified in Section~\ref{sec:choking}), and that the ambient medium is described by a power-law density profile $\rho(r) = \rhoa (r/\Rs)^{-\alpha}$.  Then, taking $t\approx \tb$ and $\Lj=E/(2\tb)$ in equation~A2 of \citet{HGN18}, we find
\begin{eqnarray}\label{eq:Rch}
\Rch & \simeq & \left[\Ns^5 C_\alpha E \rhoa^{-1} \Rs^{-\alpha} \tb^2 \theta_0^{-4}\right]^{1/(5-\alpha)},
\end{eqnarray}
where $\Ns = 0.35$ is a numerical calibration factor, $C_\alpha \equiv {(8/9\pi)(5-\alpha)^2/(3-\alpha)}$, and $\ntot$ is the total number density at $\Rs$.\footnote{Throughout the paper, we adopt solar abundance and a mean molecular weight of 0.6.}  Note that since $\rho\propto r^{-\alpha}$, the quantity $\rhoa^{-1} \Rs^{-\alpha}=(\rho r^{\alpha})^{-1}$ appearing in eq.~\ref{eq:Rch} is independent of radius; we choose to write it in this form because $\rhoa$ and $\Rs$ are both observables. We point out that the choking radius is mainly sensitive to $\tb$ and $\theta_0$, which are intrinsic properties of the central engine.  For convenience, we also introduce a dimensionless parameter
\begin{eqnarray}
\label{eq:zeta}
\zeta \equiv \frac{\Rch}{\Rs},
\end{eqnarray}
which is given by
\begin{eqnarray}\label{eq:zeta}
\zeta & = & \left[3 C_\alpha \left(\frac{E}{10^{59}\rm erg}\right) \left(\frac{\ntot}{0.1 \cmc}\right)^{-1} \right. \nonumber \\
& \times & \left. \left(\frac{\Rs}{10\rm kpc}\right)^{-5} \left(\frac{\tb}{1\rm Myr}\right)^2 \left(\frac{\theta_0}{5 \degree}\right)^{-4}  \right]^{1/(5-\alpha)}
\end{eqnarray}
We caution that our model is not equipped to handle breaks in the density profile.  Therefore, if the jet is choked inside the cold core of a cluster, the model only applies until the jet reaches the edge of the core.  The conditions for the jet to be choked inside the core of the cluster are discussed further in Section~\ref{sec:discussion}.

\begin{figure}
\begin{center}
\includegraphics[width=\columnwidth]{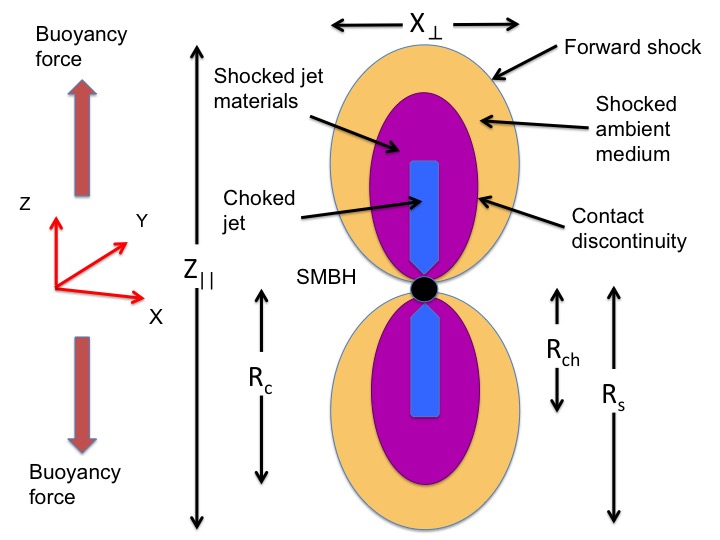} 
\caption{Schematic figure for the choked jet model during the bipolar lobes phase.} 
    \label{fig:schematic}
\end{center}
\end{figure}

After the jet is choked, the initially narrow and elongated jet-driven outflow gradually widens and transforms into a bipolar structure with two distinct lobes \citep[][hereafter I19]{INP19}, as shown in the middle panel of Fig.~\ref{fig:bubbleshape}.  In Fig.~\ref{fig:schematic}, we illustrate the geometry of the system during this bipolar lobes phase, and label the features and length scales of interest.  \footnote{Note that we define $\Rc$ and $\Rs$ as distances measured from the central black hole along the major axis, whereas \citet{TC17} defined them as distances measured from the centre of the bubbles.}

Once the forward shock has expanded over a length scale of $\sim$ tens of $\Rch$, the system enters the quasi-spherical phase (I19). In this phase, the forward shock becomes effectively spherical.  However, the evolution of the bubbles depends on whether or not buoyancy is important. In the absence of buoyancy, the bipolar bubbles also merge into a quasi-spherical shape, and the outflow resembles a spherical point explosion, as in panel 3a of Fig.~\ref{fig:bubbleshape}.  On the other hand, if buoyancy becomes important, the bipolar bubbles are pulled apart and gradually drift away from the central AGN, as shown in panel 3b of Fig.~\ref{fig:bubbleshape}.
 
\subsection{Evolutionary time-scales} \label{sec:timescales}
In order to estimate the time it takes for the outflow to become spherical, we apply the analytical results of I19,  who used an approach based on the Kompaneets approximation to derive the shape of an initially jetted, narrow outflow as a function of time after the jet is choked.  In a power-law density profile $\rho(r) \propto r^{-\alpha}$ with index $\alpha<2$, they show that the transition to quasi-spherical flow occurs on a time-scale\footnote{The time-scale $\tsp$ is defined as the time when the width of the outflow becomes 90 per cent of its height.} (I19)
\begin{eqnarray}\label{eq:tsp}
\tsp \approx 4.6 Q_\alpha \left[\frac{\rhoch \Rch^5}{E}\right]^{1/2},
\end{eqnarray} 
where $\rhoch=\rhoa \zeta^{-\alpha}$ is the density at the choking radius, and $Q_{\alpha}$ is a scaling factor relating the time when the width becomes 90 per cent of the height to the characteristic timescale $(\rhoch \Rch^5/E)^{1/2}$ (see equations 4 and 80 in I19).  For a typical ICM density profile with $\alpha=1$, $Q_\alpha \approx 450$.  In general, however, the value of $Q_\alpha$ depends strongly on $\alpha$, with $Q_{\alpha} \approx \{69, 157, 450, 3220\}$ respectively for $\alpha=\{0, 0.5,1, 1.5\}$.  The value of $Q_{\alpha}$ is also sensitive to how $\tsp$ is defined; for example, if we define $\tsp$ as the time when the width is 80 per cent of the height (instead of 90 per cent), we find $Q_\alpha \approx \{10,17,29,66\}$ for the same values of $\alpha$ (I19).  None the less, this choice of definition does not significantly affect our conclusions, because for any relevant value of $Q_\alpha$, $\tsp$ remains much larger than other timescales in the problem (see below).  By using eq.~\ref{eq:zeta} to replace $\rhoch$ and $\Rch$, eq.~\ref{eq:tsp} can be rewritten as
\begin{eqnarray}
\tsp &\approx & {\rm 11 \,Gyr}\, \zeta^{(5-\alpha)/2} \left(\frac{Q_\alpha}{450}\right) \left( \frac{\ntot}{0.1\rm cm^{-3}}\right)^{1/2} \nonumber \\
& \times & \left( \frac{\Rs}{10\rm kpc}\right)^{5/2} \left(\frac{10^{59}\rm erg}{E}\right)^{1/2}.
\end{eqnarray}

If the buoyancy force becomes important, it can remove the bubbles from the centre of the gravitational potential. 
The buoyancy time-scale for each individual bubble is estimated as $\tbuoy\sim \Rs /2\vt$,\footnote{The values of $\Rs$ and $\Rc$ defined here are roughly twice the values given in \cite{Churazov01}.} where $\vt \sim (2gV_{\rm c}/S_{\rm c}C)^{1/2}$ is the terminal velocity of the cocoon \citep{Churazov01}, $V_{\rm c}$ is the volume of
the cocoon, $S_{\rm c}$ is the cross section of the cocoon, $C=0.75$
is the drag coefficient, and $g$ is the gravitational acceleration. For a Keplerian orbit $g(R)=\vk^2(R)/R$, where $\vk$ is the Keplerian orbital velocity. If, for simplicity, we assume that the cocoon obeys spherical symmetry, we have 
\begin{eqnarray}
\tbuoy & \sim  &  \frac{0.25 \Rs^{3/2}}{ \Rc^{1/2}\vk(\Rs/2)} \nonumber \\
&\sim& 8{\rm Myr}\,\left( \frac{\Rs}{10\rm kpc}\right)^{3/2}
\left( \frac{10\rm kpc}{\Rc}\right)^{1/2} \left[\frac{300 \rm km/s}{\vk(\Rs/2)}\right],
\end{eqnarray} 
where $\vk(\Rs/2)$ is the Keplerian velocity at $\Rs/2$, which can be estimated by measuring the velocity dispersion or the mass profile in the target object. 

In a galaxy cluster, the evolution of the bubbles and the forward shock is also affected by the ambient pressure in the ICM. If the ambient pressure $\Pa$ is non-negligible, the expansion of the bubble slows down and eventually stops once it reaches pressure balance with the surrounding medium. This transition happens when the swept up energy of the ambient medium becomes comparable to the total injected energy $E$. The corresponding time-scale is\footnote{$E$ defined here is two times that in \cite{TC17}.} \citep[e.g.,][]{TC17} 
\begin{eqnarray}
t_{E}&\sim& \left(\frac{3E}{4\pi}\right)^{1/3} \frac{\rhoa^{1/2}}{\Pa^{5/6}}\sim 33{\rm Myr}\,\left(\frac{E}{10^{59}\rm erg}\right)^{1/3} \nonumber \\
&\times & \left( \frac{\ntot}{0.1\rm cm^{-3}}\right)^{1/2} \left( \frac{0.1 \kevcm}{\Pa}\right)^{5/6}.
\end{eqnarray}

The fact that $\tbuoy$ and $t_E$ are comparable is not a coincidence.  As long as the cocoon is expanding more rapidly than the terminal velocity $\vt$, the expansion overcomes buoyancy to keep the bubbles connected.  Only once the expansion speed drops below $\vt$ can the bubbles be pulled apart by buoyancy.  Now, since we consider \textit{connected} bubbles with $\Rc \sim \Rs$, $\vt$ and $\vk$ are comparable.  Additionally, it is straightforward to show that an ambient medium in virial equilibrium satisfies $\vk \sim c_{\rm s}$, where $c_{\rm s}$ is the ambient sound speed.  Thus, we see that $\vt \sim c_{\rm s}$, which means that the time $\tbuoy$ when the bubbles separate is about the same as the time $t_E$ when pressure equilibrium is achieved and the Mach number becomes order-unity.

\subsection{Choking radius and jet geometry}\label{sec:choking}

In this paper, we focus on bubbles and forward shocks which are in the bipolar lobes phase and are not strongly affected by buoyancy, as illustrated in panel 2 of Fig. \ref{fig:bubbleshape}. Bubbles uplifted by the buoyancy force are deformed and are more difficult to study due to projection effects. In general, the requirement for finding a system in the bipolar lobes phase is $\tage < \min(t_E,\tbuoy,\tsp)$.  However, our need for reliable measurements of the bubble geometry and ICM properties resulted in a sample of five well-studied, fairly evolved systems with sufficiently good spatial resolution.  In each object we consider, we find that the shocks driven by the bubbles have a low Mach number ($M<2$), indicating that they are close to reaching pressure equilibrium with the ambient medium, i.e. $\tage \sim t_E$.  Since $t_E$ and $\tbuoy$ are also comparable (as discussed above), all the systems studied here satisfy $ \tage \sim t_E \sim \tbuoy < \tsp$. 

The condition $t_E\lesssim \tsp$ provides a lower limit on $\zeta$: 
\begin{eqnarray}
\label{eq:zetamin}
\zeta^{(5-\alpha)/5} & \gtrsim & 0.1 \left(\frac{Q_\alpha}{450}\right)^{-2/5} \left( \frac{E}{10^{59} \rm erg}\right)^{1/3} \nonumber \\
& \times &  \left( \frac{0.1 \kevcm}{\Pa}\right)^{1/3}\left( \frac{10\rm kpc}{\Rs}\right).
\end{eqnarray}
Otherwise, a quasi-spherical shock is expected in the observations.  On the other hand, the jet must have been choked within the contact discontinuity, which implies $\zeta < \Rc/\Rs$.  The objects we consider have $\Rc/\Rs \simeq 0.6$--0.8.  Combining the limits on $\zeta$, we find that the choking radius is robustly constrained to be $\sim$ a few tenths of $\Rs$.  The choking time, however, is more uncertain (see Section~\ref{sec:model}).

A fundamental constraint on the jet properties comes from comparing the shock velocity upon choking, $c \betah$, to the current velocity of the forward shock.  The former velocity is expected to be faster, and therefore we have
\begin{eqnarray}
\betah \ga \dfrac{\Rs}{c \tage} = 3 \times 10^{-3} \, \left(\frac{\Rs}{10 \rm kpc}\right) \left(\frac{\tage}{10 \rm Myr}\right)^{-1}.
\label{eq:betah}
\end{eqnarray}  
This very conservative constraint simply states that the forward shock is decelerating.  This assumption holds for a constant-energy outflow with $\alpha < 3$, and for a constant-luminosity outflow with $\alpha < 2$.  For a galaxy cluster with a shallow density profile $\alpha < 2$, the assumption of a decelerating shock is reasonable.

\begin{figure}
\begin{center}
\includegraphics[width=\columnwidth]{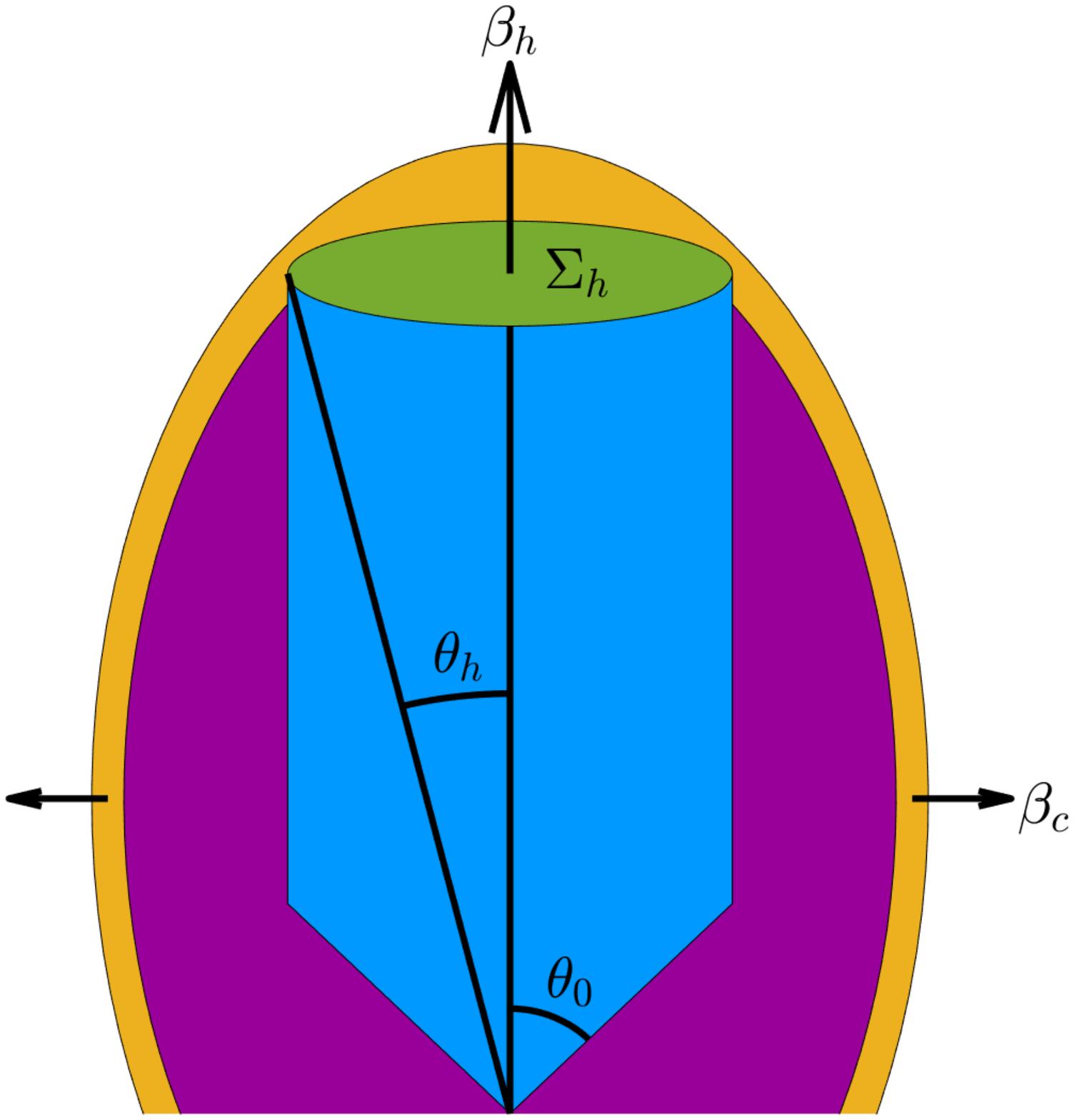} 
\caption{Geometry of the jet prior to choking, depicting the collimated jet (blue), jet head (green), cocoon (purple), and shocked ambient gas (yellow).  The head velocity $\betah$, the cocoon shock velocity $\betac$, the jet injection angle $\theta_0$, and the head's cross section $\Sigmah$ and angular size $\thetah$ are indicated.} 
 \label{fig:jetschematic}
\end{center}
\end{figure}

The lower limit on $\betah$ immediately implies an upper limit on the surface area $\Sigmah$ of the jet head, since the jet head becomes slower if the jet luminosity is spread over a larger area.  (To aid the discussion, we illustrate some relevant jet properties in Fig.~\ref{fig:jetschematic}.)  We assume that the jet head was non-relativistic, which will be checked below. Then balancing the ram pressure of the jet against that of the ambient medium leads to
\begin{eqnarray}
\Sigmah \simeq \frac{\Lj}{\rhoch \betah^2 c^3},
\label{eq:pressure_balance}
\end{eqnarray}
where $\rhoch$ is the density at $\Rch$.  Alternatively, eq.~\ref{eq:betah} can be derived by first noting that the speed of a Newtonian jet head satisfies $\betah \approx \tilde{L}^{1/2}$ \citep[B11, see also][]{Safouris08}, and then using the definition of $\tilde{L}$.   If we further assume $\Lj=E/2\tb \simeq E\betah c/2\Rch$, and apply the condition in eq.~\ref{eq:betah}, we obtain
\begin{eqnarray}
\label{eq:sigma}
\Sigmah & \simeq & \frac{E }{2\rhoch c^2 \betah \Rch} \la  0.5 \, {\rm kpc}^2 \zeta^{\alpha-1} \left(\frac{E}{10^{59} \rm erg}\right) \nonumber \\
&\times & \left( \frac{0.1\rm cm^{-3}}{\ntot}\right)\left( \frac{\tage}{10\rm Myr}\right) \left( \frac{10\rm kpc}{ \Rs}\right)^{2}.
\end{eqnarray}

Eq.~\ref{eq:sigma} assumes only that material was injected with a velocity $\approx c$ and that the reverse shock was strong while the engine was active.  It therefore applies even for spherically symmetric flows.  However, as discussed above, when the outflow was quenched its extent could not have been much less than $\Rs$, or else the forward shock would already be nearly spherical.  In order for the jet to reach the radius $\Rch$ without violating condition~\ref{eq:sigma}, the angular size of the jet head upon choking must satisfy
\begin{eqnarray}
\label{eq:theta_h}
\thetah & \simeq & \frac{(\Sigmah/\pi)^{1/2}}{\Rch} \la 1.5\degree \zeta^{(\alpha-3)/2} \left(\frac{E}{10^{59} \rm erg}\right)^{1/2} \nonumber \\
& \times & \left( \frac{0.1\rm cm^{-3}}{\ntot}\right)^{1/2} \left( \frac{\tage}{10\rm Myr}\right)^{1/2} \left( \frac{10\rm kpc}{ \Rs}\right)^{2}.
\end{eqnarray}
The upper limit on $\thetah$ suggests that the jet was either injected with a narrow opening angle or collimated by the environment.  (We note that strongly collimated jets can have $\thetah \ll \theta_0$, since in that case the jet becomes cylindrical far below the jet head.  Therefore, the upper limit on $\thetah$ given by eq.~\ref{eq:theta_h} does not necessarily imply that $\theta_0$ was small.)

An initially conical jet can be collimated via interaction with its cocoon if its luminosity satisfies (B11)
\begin{eqnarray}
\left(\frac{\Lj}{10^{45} \ergs}\right) \left(\frac{\theta_0}{5\degree}\right)^{-2/3} < 3 \times 10^6 \zeta^{2-\alpha} \left(\frac{\Rs}{10 \rm kpc}\right)^2 \left(\frac{\ntot}{0.1 \rm cm^{-3}}\right) .
\end{eqnarray}
In order for the jet head to be non-relativistic, the stricter condition 
\begin{eqnarray}
\left(\frac{\Lj}{10^{45} \ergs}\right)  \left(\frac{\theta_0}{5\degree}\right)^{-4} < 400 \zeta^{2-\alpha} \left(\frac{\Rs}{10 \rm kpc}\right)^2 \left(\frac{\ntot}{0.1 \rm cm^{-3}}\right)
\label{eq:nonrel}
\end{eqnarray}
must be met. For typical AGN parameters, we are certainly in the collimated regime, and most likely in the non-relativistic regime.  However, since the condition for a Newtonian head is sensitive to $\theta_0$, it is possible that the head was relativistic if the jet was sufficiently narrow.   Our assumption of a Newtonian head is justified as long as $\theta_0 \ga 1\degree$.

The above results suggest that the jet was collimated, but was it collimated by the cocoon (as in the middle panel of Fig.~\ref{fig:collimation}), or by the external pressure (as in the right panel of Fig.~\ref{fig:collimation})?  The requirement for collimation by the cocoon at $\Rch$ is that the pressure in the cocoon upon choking ($\Pch$) is greater than the ambient pressure ($\Pa$).   Our constraints allow us to test whether this is the case.  To do so, we first estimate the volume enclosed by the forward shock when the jet was choked as 
\begin{eqnarray}
\Vfs \simeq \frac{4 \pi}{3} \Rch^3 \left(\frac{\betac}{\betah} \right)^2,
\end{eqnarray} 
where $c \betac$ is the lateral speed of the cocoon shock at the moment of choking (see Fig.~\ref{fig:jetschematic}).  Now, if the jet is collimated by its own cocoon, the energy swept up by the forward shock is negligible compared to the energy contained in the cocoon.  Since the cocoon and shocked ambient gas are in pressure equilibrium, this implies that the cocoon takes up most of the space within the forward shock, with the shocked ambient gas only occupying a small fraction of the total volume.  This can also be seen explicitly in numerical simulations of collimated jets \citep[e.g.,][]{HGN18}. For an adiabatic index of $4/3$, the pressure in the cocoon upon choking is therefore related to the volume by $\Pch \simeq E/3\Vfs$ (B11).  Furthermore, if the ambient pressure is negligible, the sideways expansion speed of the cocoon is set by balancing $\Pch$ against the ambient ram pressure, resulting in $\Pch \simeq \rhoch c^2 \betac^2$.  
Combining the two expressions for $\Pch$, we find
\begin{eqnarray}
\Pch & \simeq & \left(\frac{E \rhoch c^2 \betah^2}{4 \pi \Rch^3} \right)^{1/2} \nonumber \\
& \ga & 0.3 \kevcm \, \zeta^{-(3+\alpha)/2} \left(\frac{E}{10^{59} \rm erg}\right)^{1/2} \nonumber \\
& \times & \left( \frac{\ntot}{0.1\rm cm^{-3}}\right)^{1/2}\left( \frac{10\rm Myr}{\tage}\right) \left( \frac{10\rm kpc}{ \Rs}\right)^{1/2}.
\end{eqnarray}
If ambient pressure were included, the cocoon would be more confined and its pressure would be even higher.  

Comparing $\Pch$ to the observed ambient pressure, we find that $\Pch \ga \Pa$, indicating that the jet was in the cocoon-collimated regime.  This is not too surprising, since all our objects have $\tage \sim t_E$, which suggests that external pressure is just now becoming important.  
 
Because the energy of the outflow is known, the lower bound on the cocoon pressure corresponds to an upper limit on the cocoon volume.  This in turn implies a lower limit on the aspect ratio upon choking, since the height $\Rch$ is restricted to a narrow range.  Using $\betac \simeq (\Pch/\rhoch c^2)^{1/2}$, we obtain
\begin{eqnarray}
\label{aspect_min}
\frac{\betah}{\betac} &  \simeq & \left(\frac{4 \pi \betah^2 \Rch^3 \rhoch c^2}{E} \right)^{1/4} \nonumber \\
& \ga & 1.3 \zeta^{(3-\alpha)/4} \left(\frac{\Rs}{10\rm kpc}\right)^{5/4} \left(\frac{\ntot}{0.1 \cmc}\right)^{1/4} \nonumber \\
& \times & \left(\frac{\tage}{10\rm Myr}\right)^{-1/2} \left(\frac{E}{10^{59}\rm erg}\right)^{-1/4}.
\end{eqnarray}
The minimum aspect ratio in eq.~\ref{aspect_min} is not very restrictive, being basically consistent with a quasi-spherical outflow.  However, we stress that condition~\ref{eq:betah} is very conservative, since in principle $\betah$ could be much higher than the currently observed shock velocity.  If $c \betah$ were higher than the current shock velocity by even a factor of two, we find $\betah/\betac \ga 1.9$ for typical parameters.  In fact, as we will discuss in the next section, constraints from the observed geometry of the forward shock suggest that the aspect ratio of the outflow upon choking was larger than 2 in some cases.  It may be difficult for a non-relativistic wind to create bubbles with such a large aspect ratio, because in that case the ram pressure applied to the reverse shock is comparable to the thermal pressure in the cocoon, and therefore the forwards and sideways expansion speeds are about the same.  In relativistic jets, on the other hand, the jet ram pressure can greatly exceed the thermal pressure in the cocoon, so achieving a large enough aspect ratio is not a problem.   

\subsection{Jet duration and opening angle}\label{sec:model}
Although our model places tight constraints on the choking radius, the time when the jet was choked is more uncertain.  We introduce a second dimensionless parameter,
\begin{eqnarray}\label{eq:eta}
\eta \equiv \frac{\tchoke}{\tage} \approx \frac{\tb}{\tage},
\end{eqnarray}
which relates the outburst duration to the current age.  Objects whose jet shut off long ago satisfy $\eta \ll 1$, whereas objects with an active jet have $\eta = 1$.  Since $\betah \simeq \Rch/c \tb$, eq. (\ref{eq:betah}) implies a fundamental upper limit of $\eta < \zeta$.  Beyond this, however, there are issues with degeneracy.  For a known injected energy $E \propto \Lj \tb$, eq.~\ref{eq:Rch} shows that the choking radius scales as  $\Rch \propto \theta_0^{-4/(5-\alpha)} \tb^{2/(5-\alpha)}$. Therefore, our model is compatible with either a wider, slower jet that was quenched relatively recently, or a narrower, faster jet that was choked long in the past.  Inserting eq.~\ref{eq:eta} into eq.~\ref{eq:zeta}, the degeneracy between $\eta$ and $\theta_0$ can be expressed as 
\begin{eqnarray}\label{eq:degen}
\eta \theta_0^{-2} &  \simeq &  7.6 C_\alpha^{-1/2} \zeta^{(5-\alpha)/2} \left(\frac{E}{10^{59}\rm erg}\right)^{-1/2} \nonumber \\
& \times & \left(\frac{\ntot}{0.1\cmc}\right)^{1/2} \left(\frac{\Rs}{10\rm kpc}\right)^{5/2} \left(\frac{\tage}{10\rm Myr}\right)^{-1}.
\end{eqnarray}
 
A possible way to break the degeneracy is to compare the relative size and shape of the X-ray cavity and the forward-shocked region.  As time goes on, the contact discontinuity moves inward relative to the forward shock, and therefore $\Rc/\Rs$ is  smaller for older outflows.  As a first approximation, we apply the point explosion model in \cite{TC17} to estimate $E$, $\tage$, and $\tb$.  The model assumes that each bubble is spherically symmetric and that the energy is injected non-relativistically. 
 
While the values of $E$ and $\tage$ estimated from the spherical model are most likely reliable, $\tb$ could be much smaller than what is inferred from the spherical model if the bubbles were inflated by a narrow, relativistic jet, instead of a wide, non-relativistic wind.  There are two reasons for this.  First, a jet will always reach a given radius faster than a spherical outflow with the same energy.    Second, a jet has $\Rc/\Rs \approx 1$ up to the moment of choking, whereas a wind could already have $\Rc/\Rs < 1$ when energy injection ends.  For these reasons, the value of $\tb$ obtained in the spherical model should be considered as an upper limit.  More detailed modelling is needed to understand how the contact discontinuity evolves in the choked jet case and improve the constraints on $\tb$.

The limits on $\zeta$ and $\eta$ can be tightened further by considering the geometry of the outflow.  For each object, we first estimate the aspect ratio of the forward shock, $z_\parallel/x_\bot$, from observations  (see Fig. \ref{fig:schematic} for the definition of $z_\parallel$ and $x_\bot$).  Due to projection effects, we can only place a lower limit on $z_\parallel/x_\bot$.  We then compare the observations to a model for the shape of the forward shock in a choked jet outflow (I19).  In the I19 model, the aspect ratio of the forward shock and the parameter $\zeta$ are computed as functions of $\eta$ using the Kompaneets approximation, for a given initial aspect ratio and injected energy (see their Figs. 10 and 12). We assume that the initial aspect ratio of the forward shock upon choking was $ \approx [12/(3-\alpha)]^{1/2} \theta_0^{-1}$, as is appropriate for a jet with a non-relativistic head (B11).  Then, for different choices of $\theta_0$, we evolve the I19 model until the aspect ratio is equal to the observed aspect ratio, and compute the corresponding value of $\zeta$ at that time.   Repeating this process for many choices of $\theta_0$ gives the minimum value of $\zeta$ needed to reproduce the observed aspect ratio of the forward shock, as a function of $\eta$.  (Models with smaller $\zeta$ would result in a smaller aspect ratio than the observed one, even if viewed edge-on.)   The lower limit on $\zeta$ inferred in this way is tighter than the one given in eq.~\ref{eq:zetamin}, particularly as $\eta$ approaches unity.

\begin{figure}
\begin{center}
\includegraphics[width=\columnwidth]{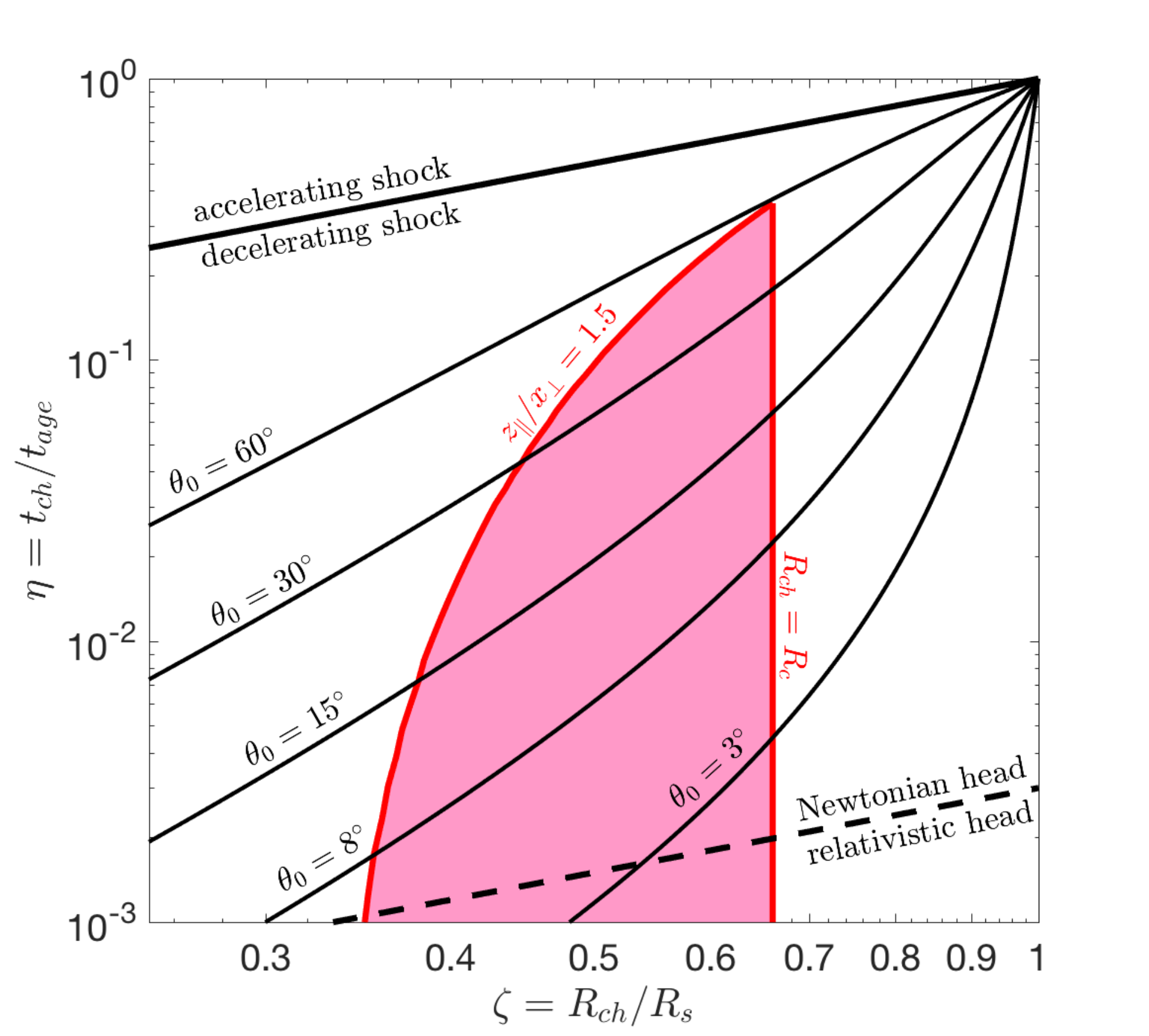} 
\caption{Constraints on $\zeta$ and $\eta$ from the geometry of Perseus.  The parameters are constrained to the pink region beneath the two red curves.  The vertical red line represents the fact that the choking radius, $\Rch$, must be less than the observed radius of the contact discontinuity, $\Rc$.  The curved red line is a curve of constant aspect ratio, $z_\parallel/r_\bot =1.5$, derived using the model of I19 (see their Fig. 10).  Above this line, the forward shock would be too spherical to explain the observations.  The dividing lines between an accelerating and decelerating shock, and between a Newtonian and relativistic jet head, are respectively indicated by the solid and dashed heavy black lines.  The thin black lines show the evolution of $\zeta$ versus $\eta$, for several choices of the jet opening angle $\theta_0$. } 
\label{fig:Perseus}
\end{center}
\end{figure}

The resulting constraints are illustrated in Fig.~\ref{fig:Perseus}, using Perseus as an example.  For this object, we estimate $z_\parallel/x_\bot \ga 1.5$ and $\Rc/\Rs \approx 0.67$, which constrains the object to lie in the pink region in the figure.  For reference, we also show several lines of constant $\theta_0$.  Although $\zeta$ is confined to a narrow range $0.33<\zeta<0.67$, we can only place upper limits on $\eta$ and $\theta_0$ due to degeneracy.  We find that $\eta < 0.36$ and $\theta_0 < 58\degree$.  The upper limit on $\tb$ is similar to the one obtained from the spherical model discussed above.   Since $E$ is well-constrained, the maximum value of $\eta$ implies a minimum value for the jet luminosity, $\Lj > 6.8 \times 10^{44}\ergs$.  

The bound on $\theta_0$ corresponds to an aspect ratio of $\ga 2.0$ when the jet was choked.  Note that this minimum value applies to the most favourable case, where the jet axis is perpendicular to the line of sight.  For a jet axis inclined by an angle $i$ with respect to the line of sight, the minimum aspect ratio becomes $2.0 \csc i$, further strengthening the case for a jetted outflow.

Because we assumed a non-relativistic jet head, our model is only valid above the dashed line in Fig.~\ref{fig:Perseus}.  This requirement can only be satisfied for $\theta_0 \ga 1 \degree$ and $\eta \ga 10^{-3}$.  However, while a Newtonian head seems likely according to eq.~(\ref{eq:nonrel}), the possibility of a relativistic head with smaller $\theta_0$ and/or $\eta$ cannot be ruled out.

Although the model permits a wide range of jet opening angles, there are a few reasons to prefer a narrower $\theta_0$.  First, the results of Section~\ref{sec:choking} suggest that the jet was strongly collimated by the cocoon, with $\thetah \la 2\degree$.  It may not be possible for wide-angle outflows to become so tightly collimated.  Models for collimation by the cocoon (e.g., B11) typically assume that $\theta_0$ is small, but when $\theta_0$ is $\sim$ tens of degrees, the physics are less well understood.   Secondly, for typical parameters, the ram pressure of the jet is much larger than the ambient pressure.  Therefore, a collimated jet is expected to produce strong shocks, at least on the axis.  The fact that we only see low Mach number shocks in every object argues against a recently quenched jet.  

\section{Results and discussion}\label{sec:discussion}

\begin{figure}
\begin{center}
\includegraphics[width=\columnwidth]{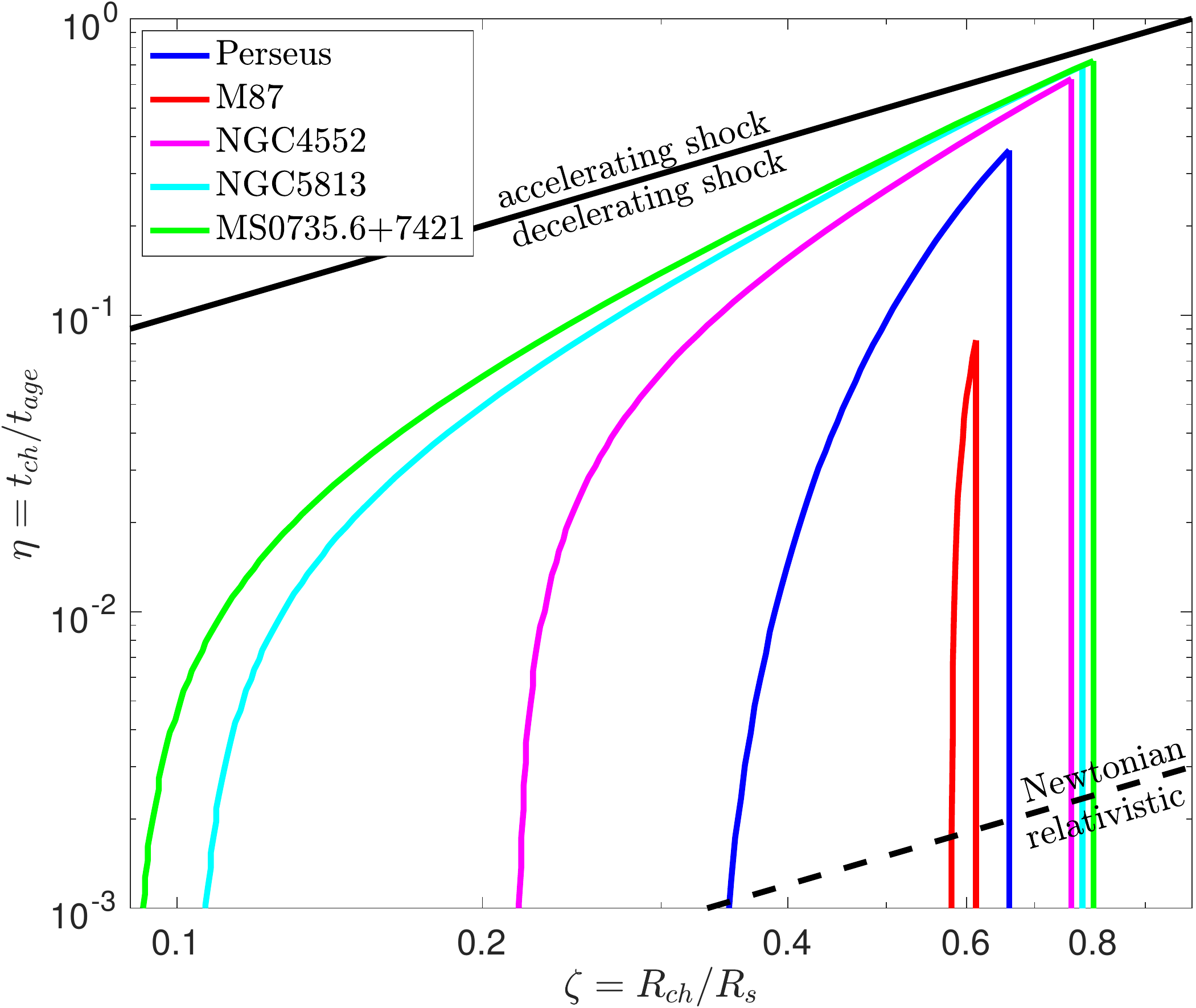} 
\caption{Constraints on $\zeta$ and $\eta$ based on the bubble geometry, as in Fig.~\ref{fig:Perseus}, but for all the objects in our sample.  For each object, the parameters are constrained to lie under the respective colored curve.}
   \label{fig:constraints}
\end{center}
\end{figure}

For each object in our sample, we repeat the process outlined in Section~\ref{sec:model}.  The resulting constraints on $\zeta$ and $\eta$ are plotted in Fig.~\ref{fig:constraints}, with each object constrained to lie under the respective coloured curve.  The upper limits on $\tb$ and $\theta_0$ and the lower limit on $\Lj$ inferred from the choked jet model are reported in Table~\ref{table:opening_angle}.  For all the objects, we find similar constraints on the choking radius, with $\zeta$ residing in the range $0.1 \la \zeta \la 0.8$.  On the other hand, for most of our sample the opening angle and jet duration are only weakly constrained.  M87, which seems to prefer a narrow jet that was choked a long time ago, is an exception.  The tighter constraints on this object are partly due to the smaller size of the bubble compared to the forward shock, and partly due to the slightly larger inferred aspect ratio (see below).

This is not the only way in which M87 is exceptional.  Although most of the objects in our sample have an aspect ratio of $\sim 1.5$, in M87 the observed aspect ratio of the shock is close to unity.  However, comparing our estimated age for this object to the time-scales from Section~\ref{sec:timescales}, we find that $\tage < \tbuoy \ll \tsp$.  If the M87 bubble was driven by a jet, then this suggests that it is still in the bipolar phase of Fig.~\ref{fig:bubbleshape}.  One explanation for the system appearing nearly spherical is that we are viewing the system along the axis.  If this is the case, we cannot directly constrain the aspect ratio from observations, but we {\it can} constrain the bubbles' {\it width}.  We therefore take a different approach to estimate $z_\parallel/x_\bot$ for this object.  For the same explosion energy and ambient density, a bipolar outflow is always more extended along the axis than a spherical blast wave.  Thus, we must have $z_\parallel > 2 R_{\rm bw}$, where $R_{\rm bw} \sim (E\tage^2/\rhoa)^{1/5}$ is the radius expected for a spherical blast wave \citep{Taylor50,Sedov59}.  For a system viewed along the axis, we also have $x_\bot = 2 \Rs$.  Therefore, the aspect ratio must satisfy $z_\parallel/x_\bot > R_{\rm bw}/\Rs \sim (E \tage^2/\rhoa \Rs^5)^{1/5}$.  Applying this method to M87, we infer a minimum aspect ratio of $z_\parallel/x_\bot \simeq 2.6$, somewhat larger than what is observed for the rest of the objects.  Another complication is that the apparent ratio of $\Rc/\Rs$ is smaller for a system observed along the axis than for a system observed edge-on.  To estimate the true value of $\Rc/\Rs$ which would be observed in the edge-on case, we assume that the bubble is roughly spherical and that its centre lies approximately halfway between the origin and the forward shock along the major axis (as is observed for the individual bubbles in our other systems).  We then find a true value of $\Rc/\Rs \simeq 0.6$ for M87, which is comparable with observations of the other four objects.

M87 is also special among the objects in our sample in that it contains both a relic bubble and a well-studied active jet.  We can therefore consider whether the jet which inflated the bubble in M87 was similar to the currently active jet.  Interestingly, the observed jet lies nearly along the line of sight \citep{BSM99,WZ09}, which is compatible with the orientation inferred from the bubble geometry.  The observed half-opening angle of the M87 jet is $\sim 30\degree$ near the base, and decreases to a steady value of about $3.5\degree$ beyond $10 \,$pc \citep{JBL99}.  The bolometric jet luminosity currently measured in M87 is $2.7\times 10^{42} \ergs$ \citep{Prieto16}, but a considerably higher mechanical luminosity of 1--3$\times 10^{44}\ergs$ is inferred from studying the kinematics of knots in the jet \citep{BB96}.  The inferred mechanical luminosity is close to the minimum value of $\sim 1 \times 10^{44} \ergs$ allowed in the bubble model, but to match this luminosity a larger jet opening angle of $\sim 20\degree$ would be required.  On the other hand, if we assume an opening angle of $\theta_0 = 3.5\degree$, as observed, we find that a much larger jet luminosity, $\sim 2 \times 10^{45} \ergs$, would be needed to inflate the observed bubble.  In conclusion, although the jet which drove the observed relic bubble had a similar orientation as the jet that we observe today, it must have been either wider or more luminous.  The difference in jet properties is consistent with the picture that the jet responsible for inflating the bubble in M87 was quenched a long time ago.  It is hard to say whether the similar orientation indicates a connection between the two episodes of jet activity, or is just a coincidence.

We now turn to the question of why the FR type I galaxies studied here appear as bipolar bubbles filled with diffuse radio emission, whereas FR type II galaxies show evidence for powerful jets and cocoons with bright radio hotspots.  We consider two possible explanations for this phenomenon.  The first is that some jets are unable to drill through the surrounding environment, and are choked to form bubbles instead.  Typical clusters have a core region with a flat density profile of $\sim r^{-1}$ extending out to $\Rcore \sim 100$\,kpc, which contains a gas mass $\Mcore \sim 10^{13}$--$10^{14}\, \msun$.  Beyond $\Rc$, the density profile breaks to $r^{-2}$ or steeper.  The time it would take the jet to break out of the dense core region is \citep{HGN18}
\begin{eqnarray}\label{eq:tbo}
\tbo & \simeq & 30\,{\rm Myr}\, \left(\frac{C_\alpha}{3-\alpha}\right)^{-1/3} \left(\frac{\Lj}{10^{45}\ergs}\right)^{-1/3} \left(\frac{\theta_0}{5\degree}\right)^{4/3} \nonumber \\
& \times & \left(\frac{\Rcore}{100\rm kpc}\right)^{2/3} \left(\frac{\Mcore}{10^{13} \msun}\right)^{1/3},
\end{eqnarray}
where as in eq.~\ref{eq:Rch} we took $\Ns=0.35$.  If the duration of jet activity satisfies $\tb > \tbo$, the jet will successfully penetrate the cluster core, but if $\tb < \tbo$ the jet will be choked before escaping.  Again assuming a total energy $E=2 \Lj \tb$ injected over a time $\tb$, equation~\ref{eq:tbo} leads to the following condition for jet choking:
\begin{eqnarray}\label{eq:chokingcondition}
\eta \theta_0^{-2} & \la & 1700 \left(\frac{C_\alpha}{3-\alpha}\right)^{-1/2} \left(\frac{E}{10^{59}\rm erg}\right)^{-1/2}  \left(\frac{\tage}{10\rm Myr}\right)^{-1} \nonumber \\
& \times & \left(\frac{\Rcore}{100\rm kpc}\right) \left(\frac{\Mcore}{10^{13} \msun}\right)^{1/2}.
\end{eqnarray}
Eq.~\ref{eq:chokingcondition} can also be derived from eq.~\ref{eq:Rch}, by using the relation $\Mcore = \int_0^{\Rcore} 4 \pi r^2 \rho(r) dr$ and applying the condition $\Rch<\Rcore$. We see that the jet can be choked either because it was too wide, or because it was too short-lived.  All the objects considered here comfortably satisfy this condition, as shown by eq.~\ref{eq:degen}.

So far, we have considered a scenario where the jet is choked because the central engine driving it is switched off.  An alternative possibility is that the jet is disrupted by the 3D magnetic kink instability, as proposed by \citet{TB16}.  For a given galaxy density and pressure profile, there is a critical jet power, above which the jets remain stable and are able to penetrate the surrounding medium, forming powerful backflows.   Below the critical power, on the other hand, the jets become kink-unstable and eventually break apart.  After disruption, the magnetic field dissolves and the outflow becomes effectively hydrodynamic.  The collimated magnetic jets are thus transformed into wide-angle hydrodynamic outflows which inflate bubbles of relativistic plasma.   This scenario differs from the model discussed in Section~\ref{sec:bubbles} in that the jet power is continuous, rather than being switched off at $\tb$.  However, the shape of the forward shock produced by a continuous, wide-angle ($\ga$ 1 radian) outflow emanating from $\Rch$ can be approximated in our model by adopting $\tb \sim \tage$ and $\Lj \sim E/2\tage$.  (In this picture, $\Rch$ refers to the radius where the jet breaks apart.)  As discussed above, continuous energy injection is disfavoured for M87, but for the other objects in our sample it is plausible that $\tb$ and $\tage$ are comparable, to within a factor of $\sim 2$ (see Fig.~\ref{fig:constraints}).

The importance of the kink instability is governed by a dimensionless stability parameter \citep{TB16}
\begin{eqnarray}\label{eq:lambda}
\Lambda(r) = \frac{2 \gammaj \thetah}{0.03} = K_\alpha \left(\frac{\Lj}{\rho(r) r^2 \gamma_j^2 c^3}\right)^{1/6} \simeq K_\alpha \left(\frac{E \theta_0^2}{2 \tb \rho(r) r^2 c^3} \right)^{1/6},
\end{eqnarray}
where $\gammaj$ is the Lorentz factor of the jet just below the head and $K_\alpha=20(2\pi/9)^{1/2}[(5-\alpha)(3-\alpha)\pi/6]^{1/3}$.  In the second equality, we applied $\Lj = E/2\tb$, and made use of the fact that for a jet collimated by the cocoon, $\gammaj \simeq \theta_0^{-1}$ (B11).  We see that the stability of the jet simply depends on the geometric quantity $\gammaj \thetah \simeq \thetah/\theta_0$.  In jets that are too long and narrow, Alfv\'en waves can circulate the jet dozens of times before the plasma can travel from the base to the jet head, which allows the instability sufficient time to grow and disrupt the jet.  In flat density profiles ($\alpha <2$), $\Lambda$ decreases continuously as the jet moves outward.  If $\Lambda$ drops below a critical value $\Lambda_{\rm crit} \sim 2$, the jet breaks apart, resulting in an FRI source \citep{TB16}. 

Therefore, if the bubbles were produced by a jet that was disrupted by the kink instability upon reaching $\Rch$, we expect to find $\Lambda(\Rch) \la \Lambda_{\rm crit} \sim 2$.  To estimate the value of $\Lambda$ at the choking radius, we use eq.~\ref{eq:Rch} to replace $\theta_0$ in eq.~\ref{eq:lambda}.  Then, taking $K_\alpha \simeq 20$, we find 
\begin{eqnarray}
\Lambda(\Rch) & \simeq & 11  \left( \frac{E}{\rhoch \Rch^3 c^2}\right)^{1/4} \simeq 0.2 \zeta^{(\alpha-3)/4} \left(\frac{E}{10^{59}\rm erg}\right)^{1/4}\nonumber \\
& \times & \left( \frac{0.1\rm cm^{-3}}{\ntot}\right)^{1/4} \left( \frac{10\rm kpc}{ \Rs}\right)^{3/4}.
\label{eq:kink}
\end{eqnarray}
Conveniently, the dependence on $\tb$ drops out, so that $\Lambda$ only depends on well-constrained quantities.
Interestingly, all the objects discussed here have $\Lambda(\Rch) \sim 0.3$, roughly consistent with the above criterion.  It therefore seems plausible that jets disrupted by the magnetic kink instability were responsible for inflating the bubbles.

\section{Conclusion} \label{sec:conclusion}

We consider relic bubbles in 5 FR type I AGN systems: Perseus, M87, MS 0735.6+7421, NGC 4552, and NGC 5813.   Motivated by the lack of spherical symmetry in most of these systems, we explore a choked jet model in which the bubbles were inflated by bipolar jets which are no longer active, and investigate how available bubble observations constrain the properties of these jets.   

Unlike previous analytical works which assumed spherical symmetry of the bubbles, we take their asymmetry into account, and show that the quasi-ellipsoidal shape of the forward shock places tight constraints on the radius $\Rch$ at which the jet was quenched.  In every object, we find that $\Rch$ is several tenths of the current radius of the bubbles.   Combining this result with the assumption of a decelerating forward shock, we show that the jet must have been tightly collimated prior to quenching, with the jet head subtending an angle of $\thetah \la 2 \degree$.  Furthermore, our results suggest that the jet was collimated by its own cocoon, and was not influenced much by the ambient pressure while it was active.

The tight limits on the choking radius place strong constraints on the allowed properties of the jet, with the jet duration $\tb$ and initial opening angle $\theta_0$ obeying eq.~\ref{eq:degen}.  This implies that if the jet was injected with a wide angle ($\sim$ tens of degrees), then it must have been active for a time comparable to the age of the system ($\sim 10$\,Myr).  On the other hand, if the jet was narrow at injection ($\sim$ a few degrees), it must have been very short-lived, lasting only $\sim 0.1$\,Myr.  Wide, short-lived jets are ruled out because they would produce a forward shock that is more spherical than what we observe.  Narrow, long-lived jets are also excluded because in that case the reverse shock would have already propagated beyond the observed location of the contact discontinuity.  

These results may have interesting implications for the FR morphological dichotomy.  We speculate that diffuse radio bubbles, like the ones considered here, originate from jets that could not traverse the cluster's core before turning off, because they were unusually wide and/or unusually short-lived.  An alternative possibility--that the bubbles were inflated by jets disrupted by the magnetic kink instability, as suggested by \citet{TB16}--also seems broadly consistent with the available data.  Since this second scenario involves a wide outflow, it can only work if the jet was disrupted on a time-scale comparable with the age of the bubbles.

The choked jet model we consider provides a time-dependent alternative to the traditional, steady-state picture of a turbulently flaring FRI jet continuously feeding its lobes \citep[e.g.,][]{Bicknell84,Bicknell95,Hardee95,Laing14}.  In this picture, internal shocks and entrainment cause the jet to decelerate and transition from laminar to turbulent flow.
This picture bears similarities to the \citet{TB16} model discussed above, except that the jet is disrupted by turbulence rather than the development of the kink instability.  In comparing our model with the \citet{TB16} scenario, we concluded that a wide-angle jet with a duration $\tb$ comparable to the current age $\tage$ is difficult to tell apart from a magnetically disrupted jet, and the same holds true in regards to the classical picture.  That is to say, in cases where it is not clear whether the X-ray cavities are relics (i.e., when $\tb \sim \tage$), we expect the choked jet model and the turbulently flaring jet model to be difficult to distinguish observationally.  However, at least in some cases, our model suggests that $\tb$ is considerably shorter than $\tage$, lending confidence to the idea that these bubbles are indeed relics of past jet activity.

An important caveat to the above discussion is that we only consider cases in which the outflow is injected at a relativistic speed.  We cannot exclude the possibility that the bubbles were inflated by quasi-spherical, non-relativistic outflows, as considered previously by, e.g., \citet{TC17} and references therein.  However, in Perseus and M87, we estimate that the aspect ratio of the outflow was $\ga 2$ when energy injection ended.  While this is not conclusive proof of a jet, it may be difficult to achieve via a non-relativistic wind.  

Among the objects we investigate, M87 serves as a useful case study, because it contains both a bubble inflated by past jet activity, and a currently active jet.  Comparing the observed jet properties to the constraints that we infer from the bubble geometry gives valuable insight into the system's history.  The shock surrounding the bubble in M87 is unique because of its apparent spherical symmetry.  However, we find that the age of the system, $\tage \sim 15$\,Myr, is much smaller than the time it would take for the forward shock to become spherical, $\tsp \sim 13$\,Gyr.  Based on this, we suggest that the bubble is not truly spherical, but rather is a bipolar structure viewed along the symmetry axis.  This idea aligns well with previous studies by \citet{BSM99} and \citet{WZ09}, who suggested that the jet axis in M87 lies nearly along the line of sight.  Due to the small size of the bubble relative to the forward shock in M87, we infer that the outburst that drove the bubbles was quenched long ago, after no more than $\sim 1.5$\,Myr of activity, in agreement with \citet{TC17}.  However, if we adopt the luminosity and opening angle of the currently observed jet, we are unable to reproduce the measured bubble properties.  We conclude that the bubble in M87 was inflated by a previous episode of jet activity that ended roughly $\sim 15$\,Myr ago, and that during that time the jet was either wider or more luminous than the jet we observe today.

Unfortunately, for the rest of the objects in our sample, we are unable to distinguish between a narrow, short-lived jet and a wider, longer-lived one.   This situation may be improved by a more complete theory.  The present model is limited in that it only considers the shape of the forward shock; a more sophisticated model that also treats the shape of the contact discontinuity in choked jet outflows is needed.  By comparing the \textit{relative} size and shape of the bubble and forward shock, the degeneracy between $\tb$ and $\theta_0$ can potentially be resolved.  Catching a system during the early evolution (i.e. during the transition from the jet phase to the bipolar lobes phase of Fig.~\ref{fig:bubbleshape}), would also go a long way towards improving our understanding of jet-inflated AGN bubbles. 

\section*{Acknowledgements}
We thank O. Bromberg and E. Churazov for helpful discussions.  
We also thank the anonymous referee for valuable comments and suggestions which improved the paper.
This research is supported by the CHE-ISF I-Core Center for Excellence in Astrophysics. 
TP is supported by an advanced ERC grant TReX. 
This work was supported in part by the Zuckerman STEM Leadership Program.
 
 \begin{table*}
 \centering
 \caption{Glossary of symbols}
 \begin{tabular}{l l}
\hline \hline
\multicolumn{2}{c}{Observables} \\ \hline 
$\Rs$ & observed radius of the forward shock along the major axis \\
$\Rc$ & observed radius of the contact discontinuity along the major axis \\
$M$ & Mach number of the forward shock \\
$z_\parallel/x_\bot$ & observed aspect ratio of the forward shock \\
$\ntot$ & total number density of the ambient medium at $r=\Rs$ \\
$\rhoa$ & mass density of the ambient medium at $r=\Rs$ \\
$\Pa$ & pressure of the ambient medium at $r=\Rs$ \\
$\alpha$ & power-law index of the ambient density profile \\
\hline
\multicolumn{2}{c}{Model parameters}  \\
\hline
$E$ & total energy injected by the jet \\
$\tage$ & age of the system \\
$\tb$ & duration of jet activity \\
$\theta_0$ & opening angle of the injected jet \\
$\Lj$ & one-sided jet luminosity, i.e. $\Lj = E/2\tb$ \\
\hline
\multicolumn{2}{c}{Evolutionary time-scales}  \\
\hline
$\tsp$ & time for the forward shock to become spherical \\
$\tbuoy$ & time-scale for buoyancy to pull the bubbles apart \\
$t_E$ & time when the outflow reaches pressure equilibrium with the ambient medium \\
\hline
\multicolumn{2}{c}{Properties of the outflow upon choking} \\
\hline
$\Rch$ & radius at which the jet is choked \\
$\tchoke$ & time at which the jet is choked \\
$\zeta$ & dimensionless ratio of $\Rch/\Rs$ \\
$\eta$ & dimensionless ratio of $\tchoke/\tage$ \\
$\rhoch$ & density of the ambient medium at $\Rch$ \\
$\betah$ & velocity of the jet head upon choking \\
$\betac$ & velocity of the cocoon shock upon choking \\
$\betah/\betac$ & estimate for the aspect ratio of the forward shock upon choking \\
$\Sigmah$ & surface area of the jet head upon choking \\
$\thetah$ & angle subtended by the jet head upon choking \\
$\Vfs$ & volume enclosed by the forward shock upon choking \\
$\Pch$ & pressure in the cocoon upon choking \\
$\Lambda(\Rch)$ & parameter determining susceptibility to the magnetic kink instability \\
 \hline \hline
 \end{tabular}
 \label{table:symbols}
 \end{table*}
 
\begin{table*}
\centering
\caption{Observed properties}
\begin{threeparttable}
\begin{tabular}{cccccccccc}
\hline\hline
Name & $\ntot(\rm cm^{-3})$ & $P(\kevcm)$& $\Rc(\rm kpc)$ & $\Rs(\rm kpc)$ & Mach number  & $\alpha$ & $z_\parallel/x_\bot$ & Reference\\
\hline \\
Perseus & 0.10 & 0.33 & 15.0 & 22.5 & 1.16 & 0.0 & 1.5 & \cite{Zhuravleva16}\\
M87 & 0.03 & 0.06 & 3.0 & 13.0 & 1.2 & 1.0& $\sim 1$& \cite{Forman17}\\
MS 0735.6+7421a & 0.01 & 0.04 & 259.0 & 320.0 &  1.3 & 1.7 & 1.4 & \cite{Van14}\\
NGC 4552 & 0.10 & 0.04 & 1.0 & 1.3 &  1.7 & 1.3 & 1.5 & \cite{Mach06}\\
NGC 5813 & 0.06 & 0.04 & 11.6 & 14.9 &  1.5 & 1.5 & 1.4 & \cite{Randall15}\\
\hline\hline
\end{tabular} 
\end{threeparttable}
\label{table:data}
\end{table*}

\begin{table*}
\centering
\caption{Derived quantities}
\begin{threeparttable}
\begin{tabular}{ccccccccc}
\hline\hline
Name & $\tage(\rm Myr)$ & $E (10^{57} \rm erg)$ &$\Rch(\rm kpc)$&$\betah/\betac$&$\Lj(10^{42} \ergs)$&$\tb(\rm Myr)$&$\theta_0$&$\Lambda(\Rch)$ \\
\hline \\
Perseus & 13.5 & 2.10e+02 & 7.5--15.0 & $\ga2.0$ & $\ga$ 680 & $\la 4.9$ & $\la $58 \degree  & 0.3--0.5  \\
M87 & 15.3 & 9.27e+00 & 7.5--8.1 & $\ga 6.7$ & $\ga 120$ & $\la 1.2$ & $\la $21\degree  & --  \\
MS 0735.6+7421a  & 97.6 & 7.71e+04 & 29--259 & $\ga 1.4$  & $\ga$ 1.9e4 & $\la 66$ & $\la90$ \degree  & 0.2--0.3  \\
NGC 4552  & 2.3 & 7.45e-02 & 0.3--1.0 & $\ga 1.6$ & $\ga$ 0.82 & $\la 1.4$ & $\la $85\degree  & 0.2--0.4  \\
NGC 5813  & 14.2 & 1.17e+01 & 1.6--11.6 & $\ga 1.5$ & $\ga$ 21 & $\la 8.7$ & $\la 90$\degree  & 0.1--0.3 \\
\hline\hline
\end{tabular} 
\end{threeparttable}
\label{table:opening_angle}
\end{table*}

\clearpage

\bsp	
\label{lastpage}
\end{document}